\documentclass[conference]{IEEEtran}
\IEEEoverridecommandlockouts

\usepackage{amssymb, amsmath}
\usepackage{amsfonts}
\usepackage{bbm}
\usepackage{mathrsfs}
\usepackage{xspace}
\usepackage{bm}

\usepackage{cite}
\usepackage{epsfig}

\newtheorem{theorem}{Theorem}

\newtheorem{proposition}{Prop.}
\newtheorem{lemma}{Lemma}

\newtheorem{assumption}{Assumption}

\newenvironment{textbmatrix}{   \setlength{\arraycolsep}{2.5pt}%
                                                                \big[\begin{matrix}}{\end{matrix}\big]%
                                                                \raisebox{0.08ex}{\vphantom{M}}}

\def\be{\begin{equation}}
\def\ee{\end{equation}}
\def\een{\nonumber \end{equation}}
\def\mat{\begin{bmatrix}}
\def\emat{\end{bmatrix}}
\def\btm{\begin{textbmatrix}}
\def\etm{\end{textbmatrix}}

\def\ba#1\ea{\begin{align}#1\end{align}}
\def\bs#1\es{\begin{split}#1\end{split}} 
\def\bg#1\eg{\begin{gather}#1\end{gather}} 
\def\bi#1\ei{\begin{itemize}#1\end{itemize}}

\newcommand{\safemath}[2]{\newcommand{#1}{\ensuremath{#2}\xspace}}

\DeclareMathOperator{\diag}{diag}                       
\DeclareMathOperator*{\argmax}{arg\;max}                
\DeclareMathOperator{\Prob}{\mathbb{P}}         

\safemath{\interior}{\mathrm{Int}}                       
\newcommand{\herm}[1]{\ensuremath{#1^{H}}}      

\safemath{\dfn}{:=}                                                     
\safemath{\dirac}{\delta}                                       

\safemath{\SNR}{\text{\sc snr}}                                 
\safemath{\No}{N_0}                                                     
\safemath{\Es}{E_s}                                                     
\safemath{\Eb}{E_b}                                                     
\safemath{\EbNo}{\frac{\Eb}{\No}}
\safemath{\EsNo}{\frac{\Es}{\No}}

\DeclareMathOperator{\CHop}{\ensuremath{\mathbb{H}}} 
\safemath{\tvir}{h_{\CHop}}                                     
\safemath{\tvtf}{L_{\CHop}}                                     
\safemath{\spf}{S_{\CHop}}                                              
\safemath{\bff}{H_{\CHop}}                                      

\safemath{\ircf}{R_{h}}                                         
\safemath{\scf}{R_{S}}                                          
\safemath{\tfcf}{R_{L}}                                         
\safemath{\bfcf}{R_{H}}                                         

\safemath{\mi}{I}                                                       
\safemath{\capacity}{C}                                         

\safemath{\normal}{\mathcal{N}}                         
\safemath{\circnorm}{\mathcal{CN}}                      
\safemath{\mchain}{\leftrightarrow}                     

\safemath{\dB}{\,\mathrm{dB}}
\safemath{\dBm}{\,\mathrm{dBm}}
\safemath{\Hz}{\,\mathrm{Hz}}
\safemath{\kHz}{\,\mathrm{kHz}}
\safemath{\MHz}{\,\mathrm{MHz}}
\safemath{\GHz}{\,\mathrm{GHz}}
\safemath{\s}{\,\mathrm{s}}
\safemath{\ms}{\,\mathrm{ms}}
\safemath{\mus}{\,\mathrm{\mu s}}
\safemath{\ns}{\,\mathrm{ns}}
\safemath{\meter}{\,\mathrm{m}}
\safemath{\mm}{\,\mathrm{mm}}
\safemath{\cm}{\,\mathrm{cm}}
\safemath{\m}{\,\mathrm{m}}
\safemath{\W}{\,\mathrm{W}}
\safemath{\J}{\,\mathrm{J}}
\safemath{\K}{\,\mathrm{K}}
\safemath{\bit}{\,\mathrm{bit}}

\safemath{\define}{\triangleq}                  

\safemath{\equivalent}{\sim}
\safemath{\distas}{\sim}                                        

\safemath{\reals}{\mathbb{R}}
\safemath{\positivereals}{\mathbb{R}^{+}}
\safemath{\integers}{\mathbb{Z}}
\safemath{\posint}{\mathbb{Z}_{+}}
\safemath{\naturals}{\mathbb{N}}
\safemath{\complexset}{\mathbb{C}}

\safemath{\setA}{\mathcal{A}}
\safemath{\setB}{\mathcal{B}}
\safemath{\setC}{\mathcal{C}}
\safemath{\setD}{\mathcal{D}}
\safemath{\setE}{\mathcal{E}}
\safemath{\setF}{\mathcal{F}}
\safemath{\setG}{\mathcal{G}}
\safemath{\setH}{\mathcal{H}}
\safemath{\setI}{\mathcal{I}}
\safemath{\setJ}{\mathcal{J}}
\safemath{\setK}{\mathcal{K}}
\safemath{\setL}{\mathcal{L}}
\safemath{\setM}{\mathcal{M}}
\safemath{\setN}{\mathcal{N}}
\safemath{\setO}{\mathcal{O}}
\safemath{\setP}{\mathcal{P}}
\safemath{\setQ}{\mathcal{Q}}
\safemath{\setR}{\mathcal{R}}
\safemath{\setS}{\mathcal{S}}
\safemath{\setT}{\mathcal{T}}
\safemath{\setU}{\mathcal{U}}
\safemath{\setV}{\mathcal{V}}
\safemath{\setW}{\mathcal{W}}
\safemath{\setX}{\mathcal{X}}
\safemath{\setY}{\mathcal{Y}}
\safemath{\setZ}{\mathcal{Z}}
\safemath{\emptySet}{\varnothing}

\safemath{\bma}{\mathbf{a}}
\safemath{\bmb}{\mathbf{b}}
\safemath{\bmc}{\mathbf{c}}
\safemath{\bmd}{\mathbf{d}}
\safemath{\bme}{\mathbf{e}}
\safemath{\bmf}{\mathbf{f}}
\safemath{\bmg}{\mathbf{g}}
\safemath{\bmh}{\mathbf{h}}
\safemath{\bmi}{\mathbf{i}}
\safemath{\bmj}{\mathbf{j}}
\safemath{\bmk}{\mathbf{k}}
\safemath{\bml}{\mathbf{l}}
\safemath{\bmm}{\mathbf{m}}
\safemath{\bmn}{\mathbf{n}}
\safemath{\bmo}{\mathbf{o}}
\safemath{\bmp}{\mathbf{p}}
\safemath{\bmq}{\mathbf{q}}
\safemath{\bmr}{\mathbf{r}}
\safemath{\bms}{\mathbf{s}}
\safemath{\bmt}{\mathbf{t}}
\safemath{\bmu}{\mathbf{u}}
\safemath{\bmv}{\mathbf{v}}
\safemath{\bmw}{\mathbf{w}}
\safemath{\bmx}{\mathbf{x}}
\safemath{\bmy}{\mathbf{y}}
\safemath{\bmz}{\mathbf{z}}

\bmdefine{\biad}{a}
\bmdefine{\bibd}{b}
\bmdefine{\bicd}{c}
\bmdefine{\bidd}{d}
\bmdefine{\bied}{e}
\bmdefine{\bifd}{f}
\bmdefine{\bigd}{g}
\bmdefine{\bihd}{h}
\bmdefine{\biid}{i}
\bmdefine{\bijd}{j}
\bmdefine{\bikd}{k}
\bmdefine{\bild}{l}
\bmdefine{\bimd}{m}
\bmdefine{\bind}{n}
\bmdefine{\biod}{o}
\bmdefine{\bipd}{p}
\bmdefine{\biqd}{q}
\bmdefine{\bird}{r}
\bmdefine{\bisd}{s}
\bmdefine{\bitd}{t}
\bmdefine{\biud}{u}
\bmdefine{\bivd}{v}
\bmdefine{\biwd}{w}
\bmdefine{\bixd}{x}
\bmdefine{\biyd}{y}
\bmdefine{\bizd}{z}

\bmdefine{\bixid}{\xi}
\bmdefine{\bilambdad}{\lambda}
\bmdefine{\bimud}{\mu}
\bmdefine{\bithetad}{\theta}
\bmdefine{\biphid}{\phi}

\safemath{\bmia}{\biad}
\safemath{\bmib}{\bibd}
\safemath{\bmic}{\bicd}
\safemath{\bmid}{\bidd}
\safemath{\bmie}{\bied}
\safemath{\bmif}{\bifd}
\safemath{\bmig}{\bigd}
\safemath{\bmih}{\bihd}
\safemath{\bmii}{\biid}
\safemath{\bmij}{\bijd}
\safemath{\bmik}{\bikd}
\safemath{\bmil}{\bild}
\safemath{\bmim}{\bimd}
\safemath{\bmin}{\bind}
\safemath{\bmio}{\biod}
\safemath{\bmip}{\bipd}
\safemath{\bmiq}{\biqd}
\safemath{\bmir}{\bird}
\safemath{\bmis}{\bisd}
\safemath{\bmit}{\bitd}
\safemath{\bmiu}{\biud}
\safemath{\bmiv}{\bivd}
\safemath{\bmiw}{\biwd}
\safemath{\bmix}{\bixd}
\safemath{\bmiy}{\biyd}
\safemath{\bmiz}{\bizd}

\safemath{\bmxi}{\bixid}
\safemath{\bmlambda}{\bilambdad}
\safemath{\bmmu}{\bimud}
\safemath{\bmtheta}{\bithetad}
\safemath{\bmphi}{\biphid}

\safemath{\bA}{\mathbf{A}}
\safemath{\bB}{\mathbf{B}}
\safemath{\bC}{\mathbf{C}}
\safemath{\bD}{\mathbf{D}}
\safemath{\bE}{\mathbf{E}}
\safemath{\bF}{\mathbf{F}}
\safemath{\bG}{\mathbf{G}}
\safemath{\bH}{\mathbf{H}}
\safemath{\bI}{\mathbf{I}}
\safemath{\bJ}{\mathbf{J}}
\safemath{\bK}{\mathbf{K}}
\safemath{\bL}{\mathbf{L}}
\safemath{\bM}{\mathbf{M}}
\safemath{\bN}{\mathbf{N}}
\safemath{\bO}{\mathbf{O}}
\safemath{\bP}{\mathbf{P}}
\safemath{\bQ}{\mathbf{Q}}
\safemath{\bR}{\mathbf{R}}
\safemath{\bS}{\mathbf{S}}
\safemath{\bT}{\mathbf{T}}
\safemath{\bU}{\mathbf{U}}
\safemath{\bV}{\mathbf{V}}
\safemath{\bW}{\mathbf{W}}
\safemath{\bX}{\mathbf{X}}
\safemath{\bY}{\mathbf{Y}}
\safemath{\bZ}{\mathbf{Z}}

\safemath{\bZero}{\mathbf{0}}

\bmdefine{\biAd}{A}
\bmdefine{\biBd}{B}
\bmdefine{\biCd}{C}
\bmdefine{\biDd}{D}
\bmdefine{\biEd}{E}
\bmdefine{\biFd}{F}
\bmdefine{\biGd}{G}
\bmdefine{\biHd}{H}
\bmdefine{\biId}{I}
\bmdefine{\biJd}{J}
\bmdefine{\biKd}{K}
\bmdefine{\biLd}{L}
\bmdefine{\biMd}{M}
\bmdefine{\biOd}{N}
\bmdefine{\biPd}{O}
\bmdefine{\biQd}{P}
\bmdefine{\biRd}{R}
\bmdefine{\biSd}{S}
\bmdefine{\biTd}{T}
\bmdefine{\biUd}{U}
\bmdefine{\biVd}{V}
\bmdefine{\biWd}{W}
\bmdefine{\biXd}{X}
\bmdefine{\biYd}{Y}
\bmdefine{\biZd}{Z}

\bmdefine{\biDelta}{\Delta}
\bmdefine{\biLambda}{\Lambda}
\bmdefine{\biPhi}{\Phi}
\bmdefine{\biSigma}{\Sigma}
\bmdefine{\biOmega}{\Omega}
\bmdefine{\biTheta}{\Theta}

\safemath{\bimA}{\biAd}
\safemath{\bimB}{\biBd}
\safemath{\bimC}{\biCd}
\safemath{\bimD}{\biDd}
\safemath{\bimE}{\biEd}
\safemath{\bimF}{\biFd}
\safemath{\bimG}{\biGd}
\safemath{\bimH}{\biHd}
\safemath{\bimI}{\biId}
\safemath{\bimJ}{\biJd}
\safemath{\bimK}{\biKd}
\safemath{\bimL}{\biLd}
\safemath{\bimM}{\biMd}
\safemath{\bimN}{\biNd}
\safemath{\bimO}{\biOd}
\safemath{\bimP}{\biPd}
\safemath{\bimQ}{\biQd}
\safemath{\bimR}{\biRd}
\safemath{\bimS}{\biSd}
\safemath{\bimT}{\biTd}
\safemath{\bimU}{\biUd}
\safemath{\bimV}{\biVd}
\safemath{\bimW}{\biWd}
\safemath{\bimX}{\biXd}
\safemath{\bimY}{\biYd}
\safemath{\bimZ}{\biZd}

\safemath{\bDelta}{\bielta}
\safemath{\bLambda}{\biLambda}
\safemath{\bPhi}{\biPhi}
\safemath{\bSigma}{\biSigma}
\safemath{\bOmega}{\biOmega}
\safemath{\bTheta}{\biTheta}

\safemath{\veca}{\bma}
\safemath{\vecb}{\bmb}
\safemath{\vecc}{\bmc}
\safemath{\vecd}{\bmd}
\safemath{\vece}{\bme}
\safemath{\vecf}{\bmf}
\safemath{\vecg}{\bmg}
\safemath{\vech}{\bmh}
\safemath{\veci}{\bmi}
\safemath{\vecj}{\bmj}
\safemath{\veck}{\bmk}
\safemath{\vecl}{\bml}
\safemath{\vecm}{\bmm}
\safemath{\vecn}{\bmn}
\safemath{\veco}{\bmo}
\safemath{\vecp}{\bmp}
\safemath{\vecq}{\bmq}
\safemath{\vecr}{\bmr}
\safemath{\vecs}{\bms}
\safemath{\vect}{\bmt}
\safemath{\vecu}{\bmu}
\safemath{\vecv}{\bmv}
\safemath{\vecw}{\bmw}
\safemath{\vecx}{\bmx}
\safemath{\vecy}{\bmy}
\safemath{\vecz}{\bmz}
\safemath{\vecZero}{\bZero}

\safemath{\vecxi}{\bmxi}
\safemath{\veclambda}{\bmlambda}
\safemath{\vecmu}{\bmmu}
\safemath{\vectheta}{\bmtheta}
\safemath{\vecphi}{\bmphi}

\safemath{\matA}{\bA}
\safemath{\matB}{\bB}
\safemath{\matC}{\bC}
\safemath{\matD}{\bD}
\safemath{\matE}{\bE}
\safemath{\matF}{\bF}
\safemath{\matG}{\bG}
\safemath{\matH}{\bH}
\safemath{\matI}{\bI}
\safemath{\matJ}{\bJ}
\safemath{\matK}{\bK}
\safemath{\matL}{\bL}
\safemath{\matM}{\bM}
\safemath{\matN}{\bN}
\safemath{\matO}{\bO}
\safemath{\matP}{\bP}
\safemath{\matQ}{\bQ}
\safemath{\matR}{\bR}
\safemath{\matS}{\bS}
\safemath{\matT}{\bT}
\safemath{\matU}{\bU}
\safemath{\matV}{\bV}
\safemath{\matW}{\bW}
\safemath{\matX}{\bX}
\safemath{\matY}{\bY}
\safemath{\matZ}{\bZ}
\safemath{\matZero}{\bZero}

\safemath{\matDelta}{\bDelta}
\safemath{\matLambda}{\bLambda}
\safemath{\matPhi}{\bPhi}
\safemath{\matSigma}{\bSigma}
\safemath{\matOmega}{\bOmega}
\safemath{\matTheta}{\bTheta}

\safemath{\matIdentity}{\matI}

\newcommand{\ut[2]}{u_{#1}(#2)}
\newcommand{\utStar[1]}{u_{#1}^*}
\newcommand{\throughput[1]}{T_{#1}}
\newcommand{\rate[1]}{R_{#1}}
\safemath{\infobits}{D}
\safemath{\totalbits}{M}
\safemath{\userno}{K}
\safemath{\userset}{\setK}
\newcommand{\power[1]}{p_{#1}}
\newcommand{\pmin[1]}{\underline{p}_{#1}}
\newcommand{\pmax[1]}{\overline{p}_{#1}}
\newcommand{\powerset[1]}{P_{#1}}
\newcommand{\powervect[1]}{\vecp_{#1}}

\newcommand{\SINR[1]}{\gamma_{#1}}
\newcommand{\SINRinf}{\bar{\gamma}^*}
\newcommand{\f[1]}{f(#1)}
\newcommand{\fPrime[1]}{f^\prime(#1)}
\safemath{\game}{G}
\safemath{\pathno}{L}

\safemath{\srake}{S_{SRake}}
\safemath{\prake}{S_{PRake}}
\safemath{\frameno}{N_f}
\safemath{\pulseno}{N_c}
\safemath{\gain}{N}
\safemath{\SP}{\text{SP}}
\safemath{\SI}{\text{SI}}
\safemath{\MAI}{\text{MAI}}
\newcommand{\channelresponse[2]}{c_{#1}(#2)}

\newcommand{\hSP[1]}{h_{#1}^{(\SP)}}
\newcommand{\hSI[1]}{h_{#1}^{(\SI)}}
\newcommand{\hMAI[1]}{h_{#1}^{(\MAI)}}

\safemath{\chiptime}{T_c}
\newcommand{\delay[1]}{\tau_{#1}}
\newcommand{\pathgain[2]}{\alpha_{#1}^{(#2)}}

\newcommand{\vecpathgain[1]}{\bmalpha_{#1}}
\newcommand{\vecrakecoeff[1]}{\bmbeta_{#1}}
\newcommand{\matpathgain[1]}{\matA_{#1}}

\newcommand{\matCoeffHifi}{\matPhi}
\newcommand{\coeffHifi[1]}{\varphi_{#1}}
\newcommand{\SIratio[1]}{\gamma_{0,{#1}}}
\newcommand{\MAIratio[1]}{\zeta_{#1}}
\newcommand{\functionGamma[1]}{\Gamma\left({#1}\right)}
\newcommand{\bestresponse[1]}{r_{#1}}

\newcommand{\vectornorm}[1]{\left|\left|{#1}\right|\right|}
\newcommand{\powerTimesHsp}{q}
\newcommand{\loadFactor}{\rho}
\newcommand{\channelgain[1]}{h_{#1}}
\newcommand{\optimumcoeff[1]}{\xi_{#1}}
\newcommand{\Po}{P_o}
\newcommand{\varq}{\sigma^2_q}
\newcommand{\meanq}{\eta_q}

\begin{document}

\title{Game-Theoretic Power Control in Impulse Radio UWB Wireless Networks}

\author{
  \authorblockN{Giacomo Bacci,\authorrefmark{1} 
    Marco Luise\authorrefmark{1} and H.~Vincent Poor\authorrefmark{2}
    \thanks{This work was supported in part by the 
      U.S. Defense Advanced Research Projects Agency under 
      Grant No.~HR0011-06-1-0052, and in part by the
      Network of Excellence in Wireless Communications NEWCOM
      of the European Commission FP6 under Contract No. 507325.}}
  \authorblockA{\authorrefmark{1} University of Pisa - 
    Dip. Ingegneria dell'Informazione - Via Caruso, 16 - 56122 Pisa, Italy\\
    Email: giacomo.bacci@iet.unipi.it; marco.luise@iet.unipi.it}
  \authorblockA{\authorrefmark{2} Princeton University - 
    Dept. of Electrical Engineering - Olden Street - 08544 Princeton, NJ, USA\\
    Email: poor@princeton.edu}
}

\maketitle

\begin{abstract}
In this paper, a game-theoretic model for studying power control for wireless data 
networks in frequency-selective multipath environments is analyzed. The uplink of 
an impulse-radio ultrawideband system is considered. The effects of self-interference 
and multiple-access interference on the performance of Rake receivers are investigated 
for synchronous systems. Focusing on energy efficiency, a noncooperative game is 
proposed in which users in the network are allowed to choose their transmit powers 
to maximize their own utilities, and the Nash equilibrium for the proposed game is 
derived. It is shown that, due to the frequency selective multipath, the noncooperative 
solution is achieved at different signal-to-interference-plus-noise ratios, respectively 
of the channel realization. A large-system analysis is performed to derive explicit 
expressions for the achieved utilities. The Pareto-optimal (cooperative) solution is 
also discussed and compared with the noncooperative approach.
\end{abstract}

\section{Introduction}\label{sec:intro}

As the demand for wireless services increases, the need for efficient resource 
allocation and interference mitigation in wireless data networks becomes more 
and more crucial. A fundamental goal of radio resource management is transmitter 
power control, which aims to allow each user to achieve the required quality of 
service (QoS) at the uplink receiver without causing unnecessary interference to 
other users in the system. Another key issue in wireless system design is energy 
consumption at user terminals, since the terminals are often battery-powered. 
Recently, game theory has been used as an effective tool to study power control 
in data networks \cite{mackenzie, goodman1, goodman2, saraydar1, 
saraydar2, feng, meshkati1, meshkati2}. In \cite{mackenzie}, the authors 
provide motivations for using game theory to study power control in 
communication systems and ad-hoc networks. In \cite{goodman1}, power 
control is modeled as a noncooperative game in which the users choose 
their transmit powers to maximize their utilities, defined as the ratio of
throughput to transmit power. In \cite{goodman2}, a network-assisted
power-control scheme is proposed to improve the overall utility of a
direct-sequence code-division multiple access (DS-CDMA) system. In
\cite{saraydar1, saraydar2}, the authors use pricing to obtain
a more efficient solution for the power control game. 
Joint network-centric and user-centric power control are discussed in
\cite{feng}. In \cite{meshkati1}, the authors propose a power control
game for multicarrier CDMA (MC-CDMA) systems, while in \cite{meshkati2}
the effects of the receiver have been considered, particularly extending the
study to multiuser detectors and multiantenna receivers.

This work considers power control in ultrawideband (UWB) systems.
UWB technology is considered to be a potential candidate for
next-generation short-range high-speed data transmission, due to
its large spreading factor (which implies large multiuser capacity)
and low power spectral density (which allows coexistence with
incumbent systems in the same frequency bands). Commonly,
impulse-radio (IR) systems, which transmit very short pulses with a
low duty cycle, are employed to implement UWB systems \cite{win}. 
In an IR system, a train of pulses is sent and the
information is usually conveyed by the position or the polarity of
the pulses, which correspond to Pulse Position Modulation (PPM) and
Binary Phase Shift Keying (BPSK), respectively. To provide
robustness against multiple access interference (MAI), each
information symbol is represented by a sequence of pulses; the
positions of the pulses within that sequence are determined by a
pseudo-random time-hopping (TH) sequence that is specific to each
user \cite{win}. In ``classical'' impulse radio, the polarity of
those pulses representing an information symbol is always the same,
whether PPM or BPSK is employed \cite{win}. Recently,
pulse-based polarity randomization was proposed \cite{nakache},
where each pulse has a random polarity code in addition to the
modulation scheme, providing additional robustness against MAI
\cite{fishler1} and helping to optimize the spectral shape according
to US Federal Communications Commission (FCC) specifications
\cite{fcc}. Due to the large bandwidth, UWB signals have a much
higher temporal resolution than conventional narrowband or wideband
signals. Hence, channel fading cannot be assumed to be flat
\cite{molisch}, and self-interference (SI) must be taken into
account \cite{gezici}. To the best of our knowledge, this paper is
the first to study the problem of radio resource allocation in a
frequency-selective multipath environment using a game-theoretic
approach. Previous work in this area has assumed flat fading
\cite{hayajneh, sun, huang}.

Our focus throughout this work is on energy efficiency. In this kind of application 
it is often more important to maximize the number of bits transmitted per Joule of 
energy consumed than to maximize throughput. We thus propose a noncooperative (distributed) 
game in which users are allowed to choose their transmit powers according to a 
utility-maximization criterion.

The remainder of this paper is organized as follows. In Sect. \ref{sec:background}, 
we provide some background for this work. The system model is given in Sect. \ref{sec:model}. 
We describe out power control game in Sect. \ref{sec:npcg} and analyze the Nash equilibrium 
for this game. In Sect. \ref{sec:ne}, we use the game-theoretic framework along with a 
large-system analysis to evaluate the performance of the system in terms of transmit powers 
and achieved utilities. The Pareto-optimal (cooperative) solution to the power control game 
is discussed in Sect. \ref{sec:pareto}, and its performance is compared with that of the 
noncooperative approach. Numerical results are discussed in Sect. \ref{sec:results}, and 
finally some conclusions are drawn in Sect. \ref{sec:conclusions}.

\section{Background}\label{sec:background}

Consider the uplink of an IR-UWB data network, where every user wishes to
locally and selfishly choose its action to maximize its
own utility function. The strategy chosen by a user affects the performance
of the other users in the network through MAI. Furthermore, since a realistic
IR-UWB transmission takes place in frequency-selective multipath channels,
the effect of SI cannot be neglected.

Game theory \cite{mackenzie} is the natural framework for modeling and
studying such interactions. To pose the power control problem as a
noncooperative game, we first need to define a utility function suitable
for measuring energy efficiency for wireless data applications. 
A tradeoff relationship is apparent to exist
between obtaining high signal-to-interference-plus-noise ratio (SINR) and 
consuming low energy. These issues can be quantified \cite{goodman1} by 
defining the utility function of the $k$th user to be the ratio of its 
throughput $\throughput[k]$ to its transmit power $\power[k]$, i.e.
\be
  \label{eq:utility}
  \ut[k]{\powervect[]}=\frac{\throughput[k]}{\power[k]},
\ee
where $\powervect[]=[\power[1],\dots,\power[\userno]]$ is the vector of
transmit powers, with \userno denoting the number of users in the network.
Throughput here refers to the net number of information bits that are received
without error per unit time (sometimes referred to as \emph{goodput}).
It can be expressed as
\be
  \label{eq:throughput}
  \throughput[k]=\frac{\infobits}{\totalbits}\rate[k] f_s(\SINR[k]),
\ee
where \infobits and \totalbits are the number of information bits and the total 
number of bits in a packet, respectively; $\rate[k]$ and $\SINR[k]$ are the 
transmission rate and the SINR for the $k$th user, respectively; and $f_s(\SINR[k])$ 
represents the packet success rate (PSR), i.e., the probability that
a packet is received without an error. Our assumption is that a packet
will be retransmitted if it has one or more bit errors. The PSR depends
on the details of the data transmission, including its modulation, coding,
and packet size. In most practical cases, however, $f_s(\SINR[k])$ is
increasing and S-shaped\footnote{An increasing function is S-shaped if
there is a point above which the function is concave, and below which
the function is convex.} (sigmoidal). For example, in the case of BPSK
TH-IR systems in multipath fading channels, $f_s(\SINR[k])$ is given by
$(1-Q(\sqrt{\SINR[k]}))^\totalbits$, where $Q(\sqrt{\SINR[k]})$ is the bit 
error rate (BER) of user $k$, with $Q(\cdot)$ denoting the complementary 
cumulative distribution function of a standard normal random variable.

\begin{figure}
  \centering
  \includegraphics[width=8.8cm]{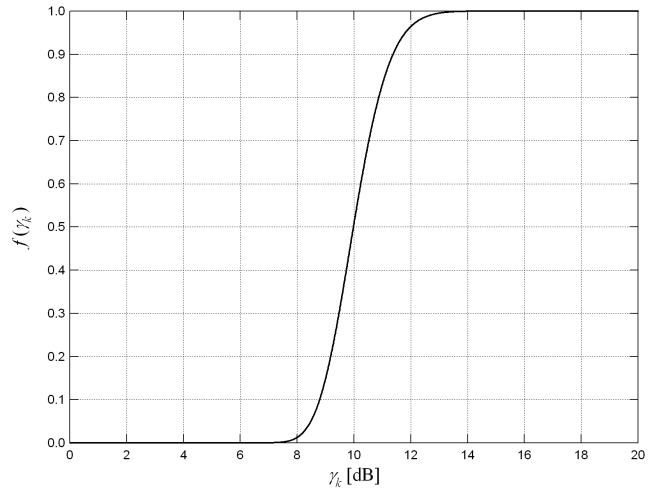}
  \caption{Typical efficiency function.}
  \label{fig:efficiency_function}
\end{figure}

To prevent the mathematical anomalies described in \cite{goodman1},
we replace PSR with an \emph{efficiency function}
$\f[{\SINR[k]}]$ when
calculating the throughput for our utility function.\footnote{Although
$\f[{\SINR[k]}]$ is introduced for analytical 
tractability, it is worth noting that typical SINRs yield negligible
differences between PSR and $\f[{\SINR[k]}]$.} 
A useful example for the efficiency function is
$\f[{\SINR[k]}]=(1-\text{e}^{-\SINR[k]/2})^\totalbits$,
which serves as a reasonable approximation to the PSR for moderate-to-large
values of \totalbits. The plot of this efficiency function is given in Fig.
\ref{fig:efficiency_function} with $\totalbits=100$
(see \cite{rodriguez} for a detailed discussion of
this efficiency function).

However, our analysis throughout this paper is valid for any efficiency
function that is increasing, S-shaped, and continuously differentiable,
with $\f[0]=0$, $\f[+\infty]=1$,
and $\fPrime[0]=0$. These assumptions are valid in many
practical systems. Furthermore, we assume that all users have the same
efficiency function. Generalization to the case where the efficiency
function is dependent on $k$ is straightforward. Note that the
throughput $\throughput[k]$ in (\ref{eq:throughput}) could also be
replaced with the Shannon capacity formula if the utility function
in (\ref{eq:utility}) is appropriately modified to ensure that
$\ut[k]{\powervect[]}=0$ when $\power[k]=0$.

Combining (\ref{eq:utility}) and (\ref{eq:throughput}), and replacing
the PSR with the efficiency function, we can write the utility function
of the $k$th user as
\be
  \label{eq:utility2}
  \ut[k]{\powervect[]}=
  \frac{\infobits}{\totalbits}\rate[k]
  \frac{\f[{\SINR[k]}]}{\power[k]}.
\ee

\begin{figure}
  \centering
  \includegraphics[width=8.8cm]{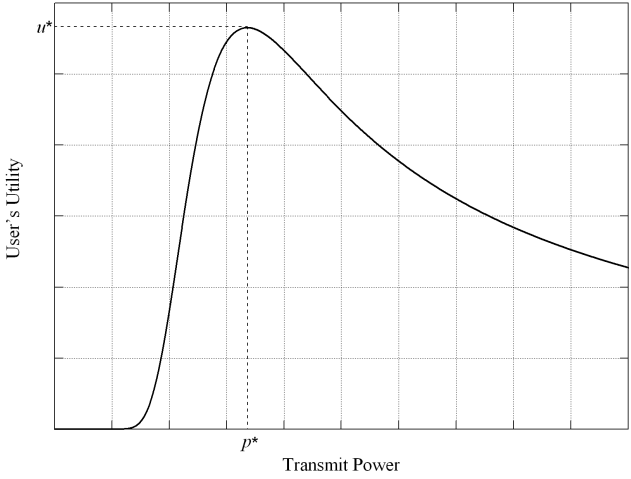}
  \caption{User's utility as a function of transmit power for fixed
    interference.}
  \label{fig:utility_shape}
\end{figure}

This utility function, which has units of bits/Joule, represents the
total number of data bits that are delivered to the destination without
an error per Joule of energy consumed, capturing the tradeoff between
throughput and battery life. Fig. \ref{fig:utility_shape} shows the 
shape of the utility function in (\ref{eq:utility2}) as a function of 
transmit power keeping other users' transmit power fixed (the meaning 
of $\power[]^*$ and $\utStar[]$ will be provided in the following).

\section{System Model}\label{sec:model}

We consider a BPSK random TH-IR system\footnote{Throughout all the paper, we analyze IR-UWB 
systems with polarity code randomization \cite{nakache}.} with \userno users in the network 
transmitting to a receiver at a common concentration point. The processing gain of the 
system is assumed to be $\gain=\frameno\cdot\pulseno$, where \frameno is the number of 
pulses that represent one information symbol, and \pulseno denotes the number of possible 
pulse positions in a frame \cite{win}. The transmission is assumed to be over 
\emph{frequency selective channels}, with the channel for user $k$ modeled as a tapped 
delay line:
\be
  \label{eq:channel}
  \channelresponse[k]{t} = 
  \sum_{l=1}^{\pathno}{\pathgain[l]{k}\delta(t-(l-1)\chiptime-\delay[k])},
\ee
where \chiptime is the duration of the transmitted UWB pulse, which is the minimum 
resolvable path interval; \pathno is the number of channel paths; and 
$\vecpathgain[k] =[\pathgain[1]{k},\dots,\pathgain[\pathno]{k}]$ and $\delay[k]$ are 
the fading coefficients and the delay of user $k$, respectively. Considering a 
chip-synchronous scenario, the symbols are misaligned by an integer multiple of the 
chip interval \chiptime: $\delay[k] = \Delta_k\chiptime$, for every $k$, where 
$\Delta_k$ is uniformly distributed in $\{0,1,\dots,\gain-1\}$. In addition we assume 
that the channel characteristics remain unchanged over a number of symbol intervals. 
This can be justified since the symbol duration in a typical application is on the 
order of tens or hundreds of nanoseconds, and the coherence time of an indoor wireless 
channel is on the order of tens of milliseconds.

Especially in indoor environments, multipath channels can have hundreds of multipath 
components due to the high resolution of UWB signals. In such cases, linear receivers 
such as matched filters (MFs), pulse-discarding receivers \cite{fishler}, and multiuser 
detectors (MUDs) \cite{verdu} cannot provide good performance, since more collisions will 
occur through multipath components. In order to mitigate the effect of multipath fading 
as much as possible, we consider a base station where \userno All-Rake (ARake) receivers 
\cite{proakis} are used.\footnote{For ease of calculation, perfect channel estimation is 
considered throughout the paper.} In particular, the ARake receiver for user $k$ is 
composed of \pathno fingers, where the vector $\vecrakecoeff[k]=\vecpathgain[k]$ 
represents the combining weights for user $k$.

The SINR of user $k$ at the output of the Rake receiver can be 
approximated (for large \frameno, typically at least 5) by \cite{gezici}
\be
  \label{eq:sinr}
  \SINR[k] = \frac{\hSP[k]\power[k]}{\displaystyle{\hSI[k]\power[k] + 
      \sum_{\substack{j=1\\j\neq k}}^{\userno}{\hMAI[kj]\power[j]} + 
      \sigma^2}},
\ee
where $\sigma^2$ is the variance of the additive white Gaussian noise (AWGN) at the 
receiver, and the gains are expressed by
\begin{align}
  \label{eq:hSP}
  \hSP[k] &= \vectornorm{\vecpathgain[k]}^2,\\
  \label{eq:hSI}
  \hSI[k] &= \frac{1}{\gain}
  \frac{\vectornorm{2\matCoeffHifi\cdot
      \herm{\matpathgain[k]}\cdot\vecpathgain[k]}^2}
       {\vectornorm{\vecpathgain[k]}^2},\\
  \label{eq:hMAI}
  \hMAI[kj] &= \frac{1}{\gain}
  \frac{\vectornorm{\herm{\matpathgain[k]}\cdot\vecpathgain[j]}^2
    + \vectornorm{\herm{\matpathgain[j]}\cdot\vecpathgain[k]}^2
  + \left|\herm{\vecpathgain[k]}\cdot\vecpathgain[j]\right|^2}
  {\vectornorm{\vecpathgain[k]}^2},
\end{align}
where the matrices
\be
  \label{eq:matrixA}
  \matpathgain[k] = 
  \begin{pmatrix}
    \pathgain[\pathno]{k}&\cdots&\cdots&\pathgain[2]{k}\\
    0&\pathgain[\pathno]{k}&\cdots&\pathgain[3]{k}\\
    \vdots&\ddots&\ddots&\vdots\\
    0&\cdots&0&\pathgain[\pathno]{k}\\
    0&\cdots&\cdots&0    
  \end{pmatrix},
\ee
and
\be
  \label{eq:matrixPhi}
  \matCoeffHifi = 
  \diag\left\{\coeffHifi[1],\cdots,\coeffHifi[\pathno-1]\right\}, 
  \qquad \coeffHifi[l]=
  \sqrt{\tfrac{\min\{\pathno-l,\pulseno\}}{\pulseno}},
\ee
have been introduced for convenience of notation.

By considering frequency selective channels, the transmit power of the $k$th 
user, $\power[k]$, does appears not only in the numerator of (\ref{eq:sinr}), 
but also in the denominator, owing to the SI due to multiple paths. In the 
following sections, we extend the approach of game theory to multipath channels, 
accounting for the SI in addition to MAI and AWGN. The problem is more challenging 
than with a single path since each user achieves a different SINR at the output 
of its Rake receiver.

\section{The Noncooperative Power Control Game}\label{sec:npcg}

In this section, we propose a noncooperative power control game (NPCG) in which 
every user seeks to maximize its own utility by choosing its transmit power. Let 
$\game = [\userset, \{\powerset[k]\}, \{\ut[k]{\powervect[]}\}]$ be the proposed 
noncooperative game where $\userset=\{1,\dots,\userno\}$ is the index set for the 
terminal users; $\powerset[k]=\left[\pmin[k], \pmax[k]\right]$ is the strategy set, 
with $\pmin[k]$ and $\pmax[k]$ denoting minimum and maximum power constraints, 
respectively; and $\ut[k]{\powervect[]}$ is the payoff function for user $k$ 
\cite{saraydar2}. Throughout this paper, we assume $\pmin[k]=0$ and 
$\pmax[k]=\pmax[]>0$ for all $k\in\userset$.

Formally, the NPCG can be expressed as
\be
  \label{eq:pc_gen}
  \max_{\power[k]\in\powerset[k]} \ut[k]{\powervect[]}=
  \max_{\power[k]\in\powerset[k]} \ut[k]{\power[k], \powervect[-k]} \qquad 
  \forall\thickspace k \in \userset,
\ee
where $\powervect[-k]$ denotes the vector of transmit powers of all terminals 
except the terminal $k$. Assuming equal transmission rate for all users, 
(\ref{eq:pc_gen}) can be rewritten as
\be
  \label{eq:pc2}
  \max_{\power[k]\in\powerset[k]} 
  \frac{\f[{\SINR[k](\power[k], \powervect[-k])}]}{\power[k]} 
  \qquad \forall\thickspace k \in \userset,
\ee
where we have explicitly shown that $\SINR[k]$ is a function of $\powervect[]$.

The solution that is most widely used for game theoretic problems is the 
\emph{Nash equilibrium}. A Nash equilibrium is a set of strategies such that no 
user can unilaterally improve its own utility. Formally, a power vector 
$\powervect[]=[\power[1],\dots,\power[\userno]]$ is a Nash equilibrium of 
$\game = [\userset, \{\powerset[k]\}, \{\ut[k]{\powervect[]}\}]$ if, for every 
$k\in\userset$, $\ut[k]{\power[k], \powervect[-k]}\ge\ut[k]{\power[k]^\prime, 
\powervect[-k]}$ for all $\power[k]^\prime\in\powerset[k]$.

\begin{theorem}\label{th:existence}
  A Nash equilibrium exists in the NPCG 
  $\game = [\userset, \{\powerset[k]\}, \{\ut[k]{\powervect[]}\}]$. Furthermore,
  the unconstrained maximization of the utility function occurs when 
  each user $k$ achieves an SINR $\SINR[k]^*$ solution of
  \be
    \label{eq:f_der}
    \SINR[k]^*\left(1-\SINR[k]^*/\SIratio[k]\right)=
    \f[{\SINR[k]^*}]/f^\prime(\SINR[k]^*),
  \ee
  where
  \be
    \label{eq:SIratio}
    \SIratio[k]=\frac{\hSP[k]}{\hSI[k]}
    =\gain\cdot
    \frac{\vectornorm{\vecpathgain[k]}^4}
         {\vectornorm{2\matCoeffHifi\cdot
             \herm{\matpathgain[k]}\cdot\vecpathgain[k]}^2}>0
  \ee
  and $f^\prime(\SINR[k])=d\f[{\SINR[k]}]/d\SINR[k]$.
\end{theorem}

\begin{lemma}\label{lm:gamma_upper_bound}
  The solution $\SINR[k]^*$ of (\ref{eq:f_der}) satisfies the condition
  \be
    \label{eq:gamma_upper_bound}
    0\le\SINR[k]^*<\SIratio[k].
  \ee
\end{lemma}
Proofs of Theorem \ref{th:existence} and Lemma \ref{lm:gamma_upper_bound} 
have been omitted because of space limitation. They can be found in \cite{bacci}.

The Nash equilibrium can be seen from another point of view. The power level chosen 
by a \emph{rational} self-optimizing user constitutes a \emph{best response} to the 
powers chosen by other players. Formally, terminal $k$'s best response 
$\bestresponse[k]: \powerset[-k]\rightarrow\powerset[k]$ is the correspondence that 
assigns to each $\powervect[-k]\in\powerset[-k]$ the set
\begin{multline}
  \label{eq:bestresponse}
  \bestresponse[k](\powervect[-k])=\{\power[k]\in\powerset[k]:
  \ut[k]{\power[k], \powervect[-k]}\ge\ut[k]{\power[k]^\prime, \powervect[-k]}\}\\
  \text{for all $\power[k]^\prime\in\powerset[k]$},
\end{multline}
where $\powerset[-k]$ is the strategy space of all users excluding user $k$.

The Nash equilibrium can be restated in a compact form: the vector $\powervect[]$ 
is a Nash equilibrium of the NPCG 
$\game = [\userset, \{\powerset[k]\}, \{\ut[k]{\powervect[]}\}]$ if and only if 
$\power[k]\in\bestresponse[k](\powervect[-k])$ for all $k\in\userset$.
\begin{proposition}
  \label{prop:bestResponse}
  Using (\ref{eq:bestresponse}), with a slight abuse of notation, user $k$'s 
  best response to a given interference vector $\powervect[-k]$ is \cite{saraydar2}
  \be
    \label{eq:prop1}
    \bestresponse[k](\powervect[-k])=\min(\pmax[],\power[k]^*),
  \ee
  where $\power[k]^*=\argmax_{\power[k]\in\positivereals}
  {\ut[k]{\power[k],\powervect[-k]}}$ is the unconstrained maximizer of the utility in 
  (\ref{eq:utility2}) (see Fig. \ref{fig:utility_shape}). Furthermore, $\power[k]^*$ is 
  unique.
\end{proposition}

\begin{proof}
  Using Theorem \ref{th:existence}, for a given interference, 
  the SINR $\SINR[k]^*$ corresponds to the transmit power $\power[k]^*$:
  \be
    \label{eq:power*}
    \power[k]^*=\frac{\SINR[k]^*\left(\sum_{j\neq k}{\hMAI[kj]\power[j]}+\sigma^2\right)}
          {\hSP[k]\left(1-\SINR[k]^*/\SIratio[k]\right)}.
  \ee

  Since $\SINR[k]^*$ is the unique maximizer of the utility, the correspondence between 
  the transmit power and the SINR must be studied. As can be verified, (\ref{eq:power*}) 
  represents the equation of a hyperbola passing through the origin, with the asymptotes 
  parallel to the Cartesian axes. In particular, the vertical asymptote is 
  $\SINR[k]^*=\SIratio[k]$. Therefore, using Lemma {\ref{lm:gamma_upper_bound}}, there 
  exists a one-to-one correspondence between the transmit power, $\power[k]^*\in[0,+\infty)$, 
  and the SINR, $\SINR[k]^*\in[0,\SIratio[k])$. Thus, the transmit power $\power[k]^*$ is 
  also unique. If $\power[k]^*\notin\powerset[k]$ for some user $k$, since it is not a 
  feasible point, then $\power[k]^*$ cannot be the best response to $\powervect[-k]$. In 
  this case, we observe that $\partial\ut[k]{\powervect[]}/\partial\power[k]\le0$ for 
  any $\SINR[k]\le\SINR[k]^*$, and hence for any $\power[k]\le\power[k]^*$. This implies 
  that the utility function is increasing in that region. Since $\pmax[]$ is the largest 
  power in the strategy space, it yields the highest utility among all $\power[k]\le\pmax[]$ 
  and thus is the best response to $\powervect[-k]$.
\end{proof}

The conclusion is that, at any equilibrium of the NPCG, a terminal either attains the 
utility maximizing SINR $\SINR[k]^*$ or it fails to do so and transmits at maximum 
power $\pmax[]$.
\begin{theorem}\label{th:uniqueness}
  The NPCG has a unique Nash equilibrium.
\end{theorem}
Proof of Theorem \ref{th:uniqueness} can be found in \cite{bacci}.

\begin{figure}
  \centering
  \includegraphics[width=8.8cm]{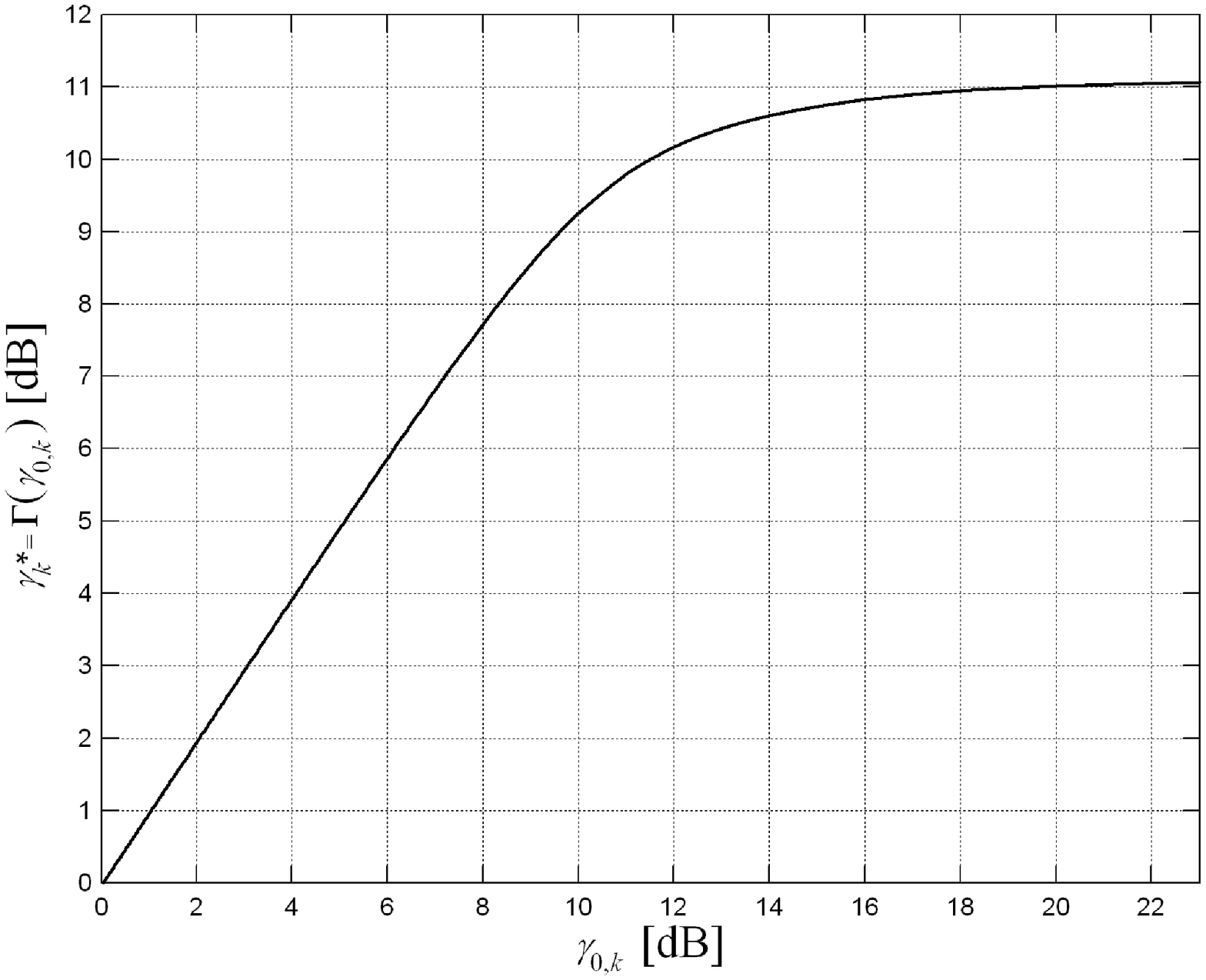}
  \caption{Shape of $\SINR[k]^*$ as a function of $\SIratio[k]$ ($\totalbits=100$).}
  \label{fig:gammaStar}
\end{figure}

\section{Analysis of the Nash equilibrium}\label{sec:ne}

In the previous section, it is seen that a Nash equilibrium for the NPCG exists 
and is unique. In the following, we study the properties of this equilibrium. It 
is worth emphasizing that, unlike previous work in this area, $\SINR[k]^*$ is 
dependent on $k$, because of the SI in (\ref{eq:sinr}). Hence, each user attains a 
different SINR. More importantly, the only term dependent on $k$ in (\ref{eq:f_der}) 
is $\SIratio[k]$, which is affected only by the channel of user $k$. This means that 
$\SINR[k]^*$ can be assumed to be constant when the channel characteristics remain 
unchanged, irrespectively of the transmit powers $\powervect[]$ and the channel 
coefficients of the other users. For convenience of notation, we can express $\SINR[k]^*$ 
as a function of $\SIratio[k]$:
\be
  \label{eq:Gamma}
  \SINR[k]^*=\functionGamma[{\SIratio[k]}].
\ee

Fig. \ref{fig:gammaStar} shows the shape of $\SINR[k]^*$ as a function of 
$\SIratio[k]$, where $\f[{\SINR[k]}]=(1-\text{e}^{-\SINR[k]/2})^\totalbits$, 
with $\totalbits=100$. Even though $\SINR[k]^*$ is shown for values of $\SIratio[k]$ 
approaching $0\,\text{dB}$, it is worth emphasizing that $\SIratio[k]>10\,\text{dB}$ 
in most practical situations.

As can be noticed, the NPCG proposed herein represents a generalization of the power 
control games discussed thoroughly in literature \cite{goodman1, goodman2, saraydar1, 
saraydar2, feng, meshkati1, meshkati2}. If $\pathno=1$, i.e. in a flat-fading scenario, 
we obtain from (\ref{eq:hSI}) and (\ref{eq:SIratio}) that $\SIratio[k]=\infty$ for all 
$k$. This implies that $\SINR[k]^*$ is the same for every $k\in\userset$, and thus it 
is possible to apply the approach proposed, e.g., in \cite{saraydar2}.

\begin{assumption}\label{ass:q}
  To simplify the analysis, let us assume the typical case of multiuser UWB systems, 
  where $\gain\gg\userno$. In addition, $\pmax[]$ is considered sufficiently large that 
  $\power[k]^*<\pmax[]$ for those users who achieve $\SINR[k]^*$. In particular, when 
  $\gain\gg\userno$, at the Nash equilibrium the following property holds:
  \be
    \label{eq:q}
    \hSP[k]\power[k]^*\simeq\powerTimesHsp>0\qquad \forall k \in\userset.
  \ee
  The heuristic derivation of (\ref{eq:q}) can be justified by SI reduction due to 
  the hypothesis $\gain\gg\userno>1$.  Using (\ref{eq:hSI}), $\SIratio[k]\gg 1$ for all 
  $k$. Hence, the noncooperative solution will be similar to that studied, e.g., in 
  \cite{meshkati2}. The validity of this assumption will be shown in Sect. \ref{sec:results} 
  through simulations.
\end{assumption}

\begin{proposition}
  \label{prop:requirement}
  A necessary and sufficient condition for a desired SINR $\SINR[k]^*$ to be achievable is
  \be 
    \label{eq:requirement}
    \SINR[k]^*\cdot\left(\SIratio[k]^{-1}+\MAIratio[k]^{-1}\right)<1\quad k\in\userset,
  \ee
  where $\SIratio[k]$ is defined in (\ref{eq:SIratio}), and 
  $\MAIratio[k]^{-1}=\sum_{j\neq k}{\hMAI[kj]/\hSP[j]}$.

  When (\ref{eq:requirement}) holds, each user can reach the optimum SINR, 
  and the minimum power solution to do so is to assign each user $k$ a transmit power
  \be
    \label{eq:minimumPower}
    \power[k]^*=\frac{1}{\hSP[k]}\cdot
    \frac{\sigma^2\SINR[k]^*}
         {1-\SINR[k]^*\cdot\left(\SIratio[k]^{-1}+\MAIratio[k]^{-1}\right)}.
  \ee
  When (\ref{eq:requirement}) does not hold, the users cannot achieve 
  $\SINR[k]^*$ simultaneously, and some of them would end up transmitting at the 
  maximum power $\pmax[]$.
\end{proposition}

\begin{proof}
  Based on Prop. \ref{prop:bestResponse}, when all users reach the Nash equilibrium, 
  their transmit powers are
    \be
    \label{eq:powerNE}
    \power[k]^*=
    \frac{\SINR[k]^*\left(\sum_{j\neq k}{\hMAI[kj]\power[j]^*}+\sigma^2\right)}
         {\hSP[k]\left(1-\SINR[k]^*/\SIratio[k]\right)}.
  \ee
  Using Assumption \ref{ass:q} in (\ref{eq:powerNE}), it is straightforward to 
  obtain:
  \be
    \label{eq:qNE}
    \powerTimesHsp\cdot
    \left[1-\SINR[k]^*\cdot\left(\SIratio[k]^{-1}+\MAIratio[k]^{-1}\right)\right]=
    \sigma^2\SINR[k]^*>0,
  \ee
  which implies $\SINR[k]^*\cdot\left(\SIratio[k]^{-1}+\MAIratio[k]^{-1}\right)<1$, 
  proving necessity. It is also straightforward to show that, if each terminal $k$ uses 
  transmit power $\power[k]^*$ as in (\ref{eq:minimumPower}), all terminals will achieve 
  the SINR requirement, finishing the proof of sufficiency. Finally, consider any other 
  joint distribution of powers and channel realizations, and let 
  $\powerTimesHsp^\prime=\inf_{k\in\userset}\left\{\hSP[k]\power[k]^*\right\}$. Then, 
  by exactly the same argument as was used in the proof of necessity,
  \be
    \label{eq:qPrime}
    \powerTimesHsp^\prime\ge
    \frac{\sigma^2\SINR[k]^*}
         {1-\SINR[k]^*\cdot\left(\SIratio[k]^{-1}+\MAIratio[k]^{-1}\right)}=
         \powerTimesHsp.
  \ee
  This means that assigning powers according to (\ref{eq:minimumPower}) does indeed 
  give the minimal power solution.
\end{proof}

Based on Prop. \ref{prop:requirement}, the amount of transmit power $\power[k]^*$ 
required to achieve the target SINR $\SINR[k]^*$ will depend not only on the gain 
$\hSP[k]$, but also on the SI term $\hSI[k]$ (through $\SIratio[k]$) and the 
interferers $\hMAI[kj]$ (through $\MAIratio[k]$). To derive some quantitative 
results independent of SI and MAI terms, it is possible to resort to a large 
systems analysis. 

\begin{assumption}\label{ass:approximation}
  Consider an exponential decaying averaged Power Delay Profile (aPDP) for the channel 
  coefficients \cite{rappaport}. Let us assume a network where $\pulseno\gg1$, 
  $\pathno\gg1$, and $\sqrt[\pulseno]{\pathno}/\pulseno=\loadFactor$, with 
  $\loadFactor\ll1$. If an ARake receiver is used,
  \be
    \label{eq:interferenceLSA}
    \left(\SIratio[k]^{-1}+\MAIratio[k]^{-1}\right) \simeq 
    \loadFactor\cdot\left(\userno-1\right)/\frameno.
  \ee
\end{assumption}

The accuracy of this approximation will be verified in Sect. \ref{sec:results} 
using simulations. 

\begin{proposition}\label{prop:sinr-balancing}
  When the hypotheses of Assumption \ref{ass:approximation} hold, using 
  (\ref{eq:requirement}) and (\ref{eq:interferenceLSA}), a necessary and sufficient 
  condition for the target SINR $\SINR[k]^*$ to be achievable is\footnote{In order 
    for the analysis to be consistent, and also considering regulations by the 
    FCC \cite{fcc}, it is worth noting that $\frameno$ could not be smaller than a 
    certain threshold ($\frameno\ge5$).}
  \be
    \label{eq:requirementLSA}
    \frameno\ge\lceil\SINRinf\cdot\loadFactor\left(\userno-1\right)\rceil,
  \ee
  where $\lceil\cdot\rceil$ is the ceiling operator, and 
  $\SINRinf=\functionGamma[\infty]$.

  In addition, the desired SINR $\SINR[k]^*$ approaches $\SINRinf$ for every user, 
  thus leading to a nearly SINR-balancing scenario.
\end{proposition}

Based on Prop. \ref{prop:sinr-balancing}, it is possible to provide expressions for 
users' transmit powers $\power[k]^*$ and utilities $\utStar[k]$ at the Nash equilibrium
(see Fig. \ref{fig:utility_shape}),
which are independent of the channel realizations of the other users and of the 
SI:\footnote{Of course, the amount of transmit power $\power[k]^*$ needed to achieve 
$\SINR[k]^*$ is dependent on the channel realization of user $k$.}
\begin{align}
  \label{eq:power*LSA}
  \power[k]^* &\simeq\frac{1}{\hSP[k]}\cdot\frac{\sigma^2\SINRinf}
  {1-\SINRinf\cdot\loadFactor\cdot\left(\userno-1\right)/\frameno},\\
  \label{eq:utilityLSA}
  \utStar[k] &\simeq\hSP[k]\cdot\frac{\infobits}{\totalbits}\rate[k]
  \frac{\f[{\SINRinf}]
    \left[1-\SINRinf\cdot\loadFactor\cdot\left(\userno-1\right)/\frameno\right]}
       {\sigma^2\SINRinf}.
\end{align}

The validity of these claims will also be confirmed through simulations in 
Sect. \ref{sec:results}.

\section{Social Optimum}\label{sec:pareto}
The solution to the power control game is said to be Pareto-optimal if there 
exists no other power allocation $\powervect[]$ for which one or more users can 
improve their utilities without reducing the utility of any of the other users. 
It can be shown that the Nash equilibrium presented in the previous section is 
not Pareto-optimal. This means that it is possible to improve the utility of one 
or more users without harming other users. On the other hand, it can be shown 
that the solution to the following social problem gives the Pareto-optimal frontier 
\cite{meshkati2}:
\be
  \label{eq:optimum1}
  \powervect[opt]= \argmax_{\powervect[]} \sum_{k=1}^{\userno}
  {\optimumcoeff[k] \ut[k]{\powervect[]}}
\ee
for $\optimumcoeff[k]\in\positivereals$. Pareto-optimal solutions are, in general, 
difficult to obtain. Here, we conjecture that the Pareto-optimal solution occurs 
when all users achieve the same SINRs, $\SINR[opt]$. This approach is chosen not 
only because SINR balancing ensures fairness among users in terms of throughput 
and delay \cite{meshkati2}, but also because, for large systems, the Nash equilibrium 
is achieved when all SINRs are similar. We also consider the hypothesis 
$\optimumcoeff[1]=\dots=\optimumcoeff[\userno]=1$, suitable for a scenario without 
priority classes. Hence, (\ref{eq:optimum1}) can be written as
\be
  \label{eq:optimum2}
  \powervect[opt]= \argmax_{\powervect[]} \f[{\SINR[]}]\sum_{k=1}^{\userno}
  {\frac{1}{\power[k]}}.
\ee
In a network where Assumptions \ref{ass:q} and \ref{ass:approximation} hold, at 
the Nash equilibrium all users achieve a certain output SINR $\SINR[]$ with 
$\hSP[k]\power[k]\simeq\powerTimesHsp\left(\SINR[]\right)$, where
\be
  \label{eq:optimumq}
  \powerTimesHsp\left(\SINR[]\right)=
  \left(\frameno\sigma^2\SINR[]\right)/
       \left[\frameno-\SINR[]\loadFactor\left(\userno-1\right)\right],
\ee
with $\loadFactor=\sqrt[\pulseno]{\pathno}/\pulseno$. Therefore, (\ref{eq:optimum2}) 
can be expressed as
\be
  \label{eq:optimum3}
  \SINR[opt]\simeq\argmax_{\SINR[]} \frac{\f[{\SINR[]}]}
       {\powerTimesHsp\left(\SINR[]\right)}
  \sum_{k=1}^{\userno}{\hSP[k]},
\ee
since there exists a one-to-one correspondence between $\SINR[]$ and $\powervect[]$. 
It should be noted that, while the maximizations in (\ref{eq:pc2}) consider no 
cooperation among users, (\ref{eq:optimum2}) assumes that users cooperate in 
choosing their transmit powers. That means that the relationship between the 
user's SINR and transmit power will be different from that in the noncooperative case.

\begin{proposition}\label{prop:so}
  In a network where $\pulseno\gg1$, $\pathno\gg1$, and $\gain\gg\userno$, the 
  Nash equilibrium approaches the Pareto-optimal solution.
\end{proposition}

\begin{proof}\label{pr:so}
  The solution $\SINR[opt]$ to (\ref{eq:optimum3}) must satisfy the condition 
  $(d\left(\f[{\SINR[]}]/\powerTimesHsp\left(\SINR[]\right)\right)/
  d\SINR[])|_{\SINR[]=\SINR[opt]}=0$. Using this fact, combined with 
  (\ref{eq:optimumq}), gives us the equation that must be satisfied by the 
  solution of the maximization problem in (\ref{eq:optimum3}):
  \be
    \label{eq:optimumf}
    f^\prime({\SINR[opt]})\SINR[opt]\left[1-\SINR[opt]\loadFactor\left(\userno-
      1\right)/\frameno\right]=\f[{\SINR[opt]}].
  \ee 
  We see from (\ref{eq:optimumf}) that the Pareto-optimal solution differs from 
  the solution (\ref{eq:f_der}) of the noncooperative utility-maximizing method, 
  since (\ref{eq:optimumf}) also takes into account the contribution of the 
  interferers. In particular,
  \be
    \label{eq:Gamma2}
    \SINR[opt]=\functionGamma[{\frac{\frameno}{\loadFactor\left(\userno-1\right)}}].
  \ee

  Since the function $\functionGamma[\cdot]$ is increasing with its argument for 
  any S-shaped $\f[{\SINR[]}]$ (as can also be seen in Fig. \ref{fig:gammaStar}), 
  and since $\frameno/\left[\loadFactor\left(\userno-1\right)\right]\le\SIratio[k]$ 
  for all $k$ (from (\ref{eq:interferenceLSA})),
  \be
    \label{eq:NEvsSO}
    \SINR[opt]\le\SINR[]\le\SINRinf, 
  \ee
  due to (\ref{eq:Gamma}) and (\ref{eq:Gamma2}). On the other hand, typical values of 
  $\pathno$ and $\pulseno$ lead to $\loadFactor\ll1$. Thus, 
  $\SINR[opt]\rightarrow\SINRinf$. From (\ref{eq:NEvsSO}), it is apparent that 
  $\SINR[]\rightarrow\SINRinf$ as well. This means that, in almost all practical 
  scenarios, the target SINR for the noncooperative game, $\SINR[]$, is close to the 
  target SINR for the Pareto-optimal solution, $\SINR[opt]$. Consequently, the average 
  utility provided by the Nash equilibrium is close to the one achieved according to 
  the Pareto-optimal solution.
\end{proof}

The validity of the above claims will be verified in Sect. \ref{sec:results} using 
simulations.

\begin{table}
  \renewcommand{\arraystretch}{1.3}
  \caption{Ratio $\varq/\meanq^2$ for different network parameters.}
  \label{tab:q}
  \centering
  \begin{tabular}{c|c|c|c|c|}
    & \multicolumn{4}{|c|}{$(\pathno, \userno)$}\\
    \hline
    $(\pulseno, \frameno)$ & \it{(20,8)} & \it{(20,16)} & \it{(50,8)} & \it{(50,16)} \\
    \hline
    \it{(30,10)} & 9.4E-4 & 3.2E-3 & 4.8E-4 & 1.7E-3 \\
    \hline
    \it{(30,50)} & 2.9E-5 & 6.4E-5 & 1.6E-5 & 3.4E-5 \\
    \hline
    \it{(50,10)} & 2.9E-4 & 6.8E-4 & 1.5E-4 & 3.7E-4 \\
    \hline
    \it{(50,50)} & 1.0E-5 & 2.2E-5 & 0.6E-5 & 1.2E-5 \\
    \hline
    \it{(100,10)} & 6.7E-5 & 1.5E-4 & 3.7E-5 & 7.8E-5 \\
    \hline
    \it{(100,50)} & 0.3E-5 & 0.6E-5 & 0.1E-5 & 0.3E-5 \\
    \hline
  \end{tabular}
\end{table}

\begin{table}
  \renewcommand{\arraystretch}{1.3}
  \caption{Normalized mean squared error of the approximation (\ref{eq:interferenceLSA}).}
  \label{tab:approximation}
  \centering
  \begin{tabular}{c|c|c|c|c|}
    & \multicolumn{4}{|c|}{$(\frameno, \userno)$}\\
    \hline
    $\loadFactor$ & \it{(20,8)} & \it{(20,12)} & \it{(50,8)} & \it{(50,12)} \\
    \hline
    \it{0.01} & 0.004 & 0.003 & 0.004 & 0.003 \\
    \hline
    \it{0.02} & 0.004 & 0.004 & 0.004 & 0.005 \\
    \hline
    \it{0.03} & 0.004 & 0.005 & 0.004 & 0.005 \\
    \hline
    \it{0.04} & 0.005 & 0.007 & 0.005 & 0.007 \\
    \hline
    \it{0.05} & 0.006 & 0.009 & 0.006 & 0.009 \\
    \hline
  \end{tabular}
\end{table}

\section{Numerical Results}\label{sec:results}

In this section, we discuss numerical results for the analysis presented in the previous 
sections. We assume that each packet contains $100\,\text{b}$ of information and no 
overhead (i.e., $\infobits=\totalbits=100$). The transmission rate is 
$\rate[]=100\,\text{kb/s}$, the thermal noise power is $\sigma^2=5 \times 10^{-16}\,\text{W}$, 
and $\pmax[]=1\,\mu\text{W}$. We use the efficiency function 
$\f[{\SINR[k]}]=(1-\text{e}^{-\SINR[k]/2})^\totalbits$, which serves as a reasonable 
approximation to the PSR for moderate-to-large values of $\totalbits$. Using 
$\totalbits=100$, $\SINRinf=\functionGamma[\infty]=11.1\,\text{dB}$. To model the UWB 
scenario, channel gains are simulated following \cite{cassioli} . The distance between 
users and base station is assumed to be uniformly distributed between $3$ and $20\,\text{m}$.

Before showing the numerical results for both the noncooperative and the cooperative 
approaches, some simulations are provided to verify the validity of Assumptions 
\ref{ass:q} and \ref{ass:approximation} introduced in Sect. \ref{sec:ne}. Table 
\ref{tab:q} reports the ratio $\varq/\meanq^2$ of the variance $\varq$ to the squared 
mean value $\meanq^2$ of $\powerTimesHsp=\hSP[k]\power[k]^*$, obtained by averaging 
$10\,000$ realizations of channel coefficients for different values of network 
parameters using ARake receivers. We can see that, when the processing gain is much 
greater than the number of users, $\varq/\meanq^2\ll1$. Hence, (\ref{eq:q}) can be 
used to carry out the theoretical analysis of the Nash equilibrium.

Table \ref{tab:approximation} shows the accuracy of the approximation 
(\ref{eq:interferenceLSA}) for the term $\left(\SIratio[k]^{-1}+\MAIratio[k]^{-1}\right)$, 
where $10\,000$ random realizations of fading coefficients are employed.\footnote{It 
is worth emphasizing that the channel model proposed in \cite{cassioli} 
fulfills the hypothesis of exponential decay of the aPDP.} Simulations are performed 
for different values of $\userno$, $\frameno$, and $\loadFactor$, using ARake 
receivers. In the right column, the ratio of the mean squared error (mse) of the 
estimation to the squared approximation $\loadFactor\cdot(\userno-1)/\frameno$ is 
reported. Like before, it can be noticed that, in the worst case, such ratio is 
smaller than $10^{-2}$, validating (\ref{eq:interferenceLSA}).

\begin{figure}
  \centering
  \includegraphics[width=8.8cm]{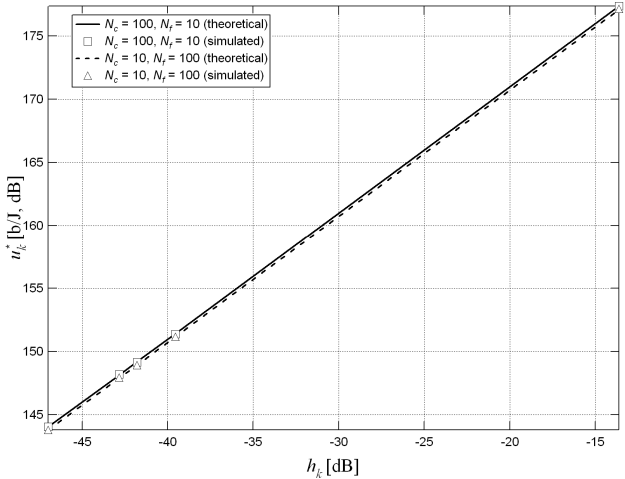}
  \caption{Utility versus channel gain at the Nash equilibrium for different ratios $\pulseno/\frameno$.}
  \label{fig:utilityNfixed}
\end{figure}

Fig. \ref{fig:utilityNfixed} shows the utility as a function of the channel gain 
$\channelgain[k]=\vectornorm{\vecpathgain[k]}^2$
when $\gain$ is constant, but the ratio $\pulseno/\frameno$ is 
variable. The results have been obtained for a network with $\userno=5$ users, 
$\pathno=50$ channel paths and $\gain=1000$, using ARake receivers 
at the base station. The lines represent the theoretical values provided by 
(\ref{eq:utilityLSA}), while the markers report the simulation results. The solid line 
corresponds to $\pulseno/\frameno=10$, and the dashed line shows $\pulseno/\frameno=0.1$. 
We can see that the simulations match closely with the theoretical results. In 
addition, as expected, higher $\pulseno/\frameno$ ratios (and thus higher $\pulseno$, 
when $\gain$ is fixed), correspond to higher utility, since $\utStar[k]/\channelgain[k]$ 
is proportional to $1-\sqrt[\pulseno]{\pathno}\cdot(\userno-1)/\gain$, which increases 
as $\pulseno$ increases. This result complies with theoretical analysis of UWB 
systems \cite{gezici}, since, for a fixed total processing gain $\gain$, increasing 
the number of chips per frame, $\pulseno$, will decrease the effects of SI, while 
the dependency of the expressions on the MAI remains unchanged. Hence, a system with 
a higher $\pulseno$ achieves better performance.

\begin{figure}
  \centering
  \includegraphics[width=8.8cm]{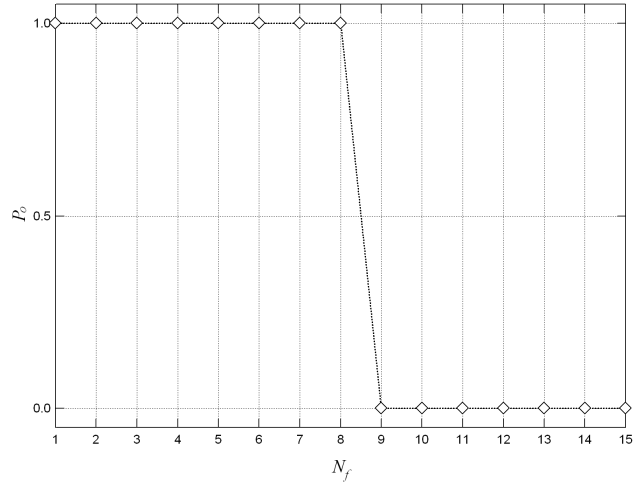}
  \caption{Probability of having at least one user transmitting at maximum power versus the number of frames.}
  \label{fig:Nfmin}
\end{figure}

Fig. \ref{fig:Nfmin} shows the probability $\Po$ of having at least one user 
transmitting at the maximum power, i.e., 
$\Po=\Prob\{\max_k\power[k]=\pmax[]=1\,\mu\text{W}\}$, as a function of the number 
of frames $\frameno$. We consider $10\,000$ realizations of the channel gains, using 
a network with ARake receivers at the base station, $\userno=32$ users, $\pulseno=50$, 
and $\pathno=100$ (thus $\loadFactor\simeq 0.0219$). In (\ref{eq:requirementLSA}), 
it has been shown that the minimum value of $\frameno$ that allow all $\userno$ users 
to achieve the optimum SINRs is 
$\frameno=\lceil\SINRinf\cdot\loadFactor\left(\userno-1\right)\rceil=\lceil 8.803 
\rceil = 9$. Simulations thus agree with the analytical results of Sect. \ref{sec:ne}.

\begin{figure}
  \centering
  \includegraphics[width=8.8cm]{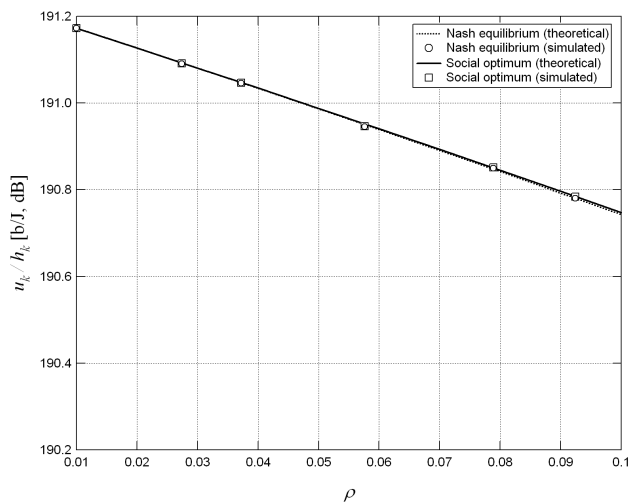}
  \caption{Comparison of the normalized utility versus load factor for the noncooperative 
    and Pareto-optimal solutions.}
  \label{fig:utilityNEvsSO}
\end{figure}

We now analyze the performance of the system when using a Pareto-optimal solution 
instead of the Nash equilibrium. Fig. \ref{fig:utilityNEvsSO} shows the normalized 
utility $u_k/\channelgain[k]$ as a function of the load factor $\loadFactor$. We 
consider a network with $\userno=5$ users, $\frameno=50$ frames and ARake receivers at 
the base station. The lines represent theoretical values of Nash equilibrium (dotted line), 
using (\ref{eq:utilityLSA}), and of the social optimum solution (solid line), using 
(\ref{eq:utilityLSA}) again, but substituting $\SINRinf$ with the numerical solution 
of (\ref{eq:optimumf}), $\SINR[opt]$. The markers correspond to the simulation results. 
In particular, the circles represent the averaged solution of the NPCG iterative algorithm, 
while the square markers show averaged numerical results (through a complete search) of the 
maximization (\ref{eq:optimum1}), with $\optimumcoeff[k]=1$. As stated in Sect. 
\ref{sec:pareto}, the difference between the noncooperative approach and the 
Pareto-optimal solution is not significant, especially for lower values of the load 
factor $\loadFactor$. Fig. \ref{fig:gammaNEvsSO} compares the target SINRs of the 
noncooperative solutions with the target SINRs of the Pareto-optimal solutions. As before, 
the lines correspond to the theoretical values (dashed line for the noncooperative solution, 
solid line for the social optimum solution), while the markers represent the simulation 
results (circles for the noncooperative solutions, square markers for the Pareto-optimal 
solution). It is seen that, in both cases, the average target SINRs for the Nash 
equilibrium, $\SINR[]$, and for the social optimum solution, $\SINR[opt]$, are very close 
to $\SINRinf$, as shown in Sect. \ref{sec:pareto}.

\section{Conclusion and Perspectives}\label{sec:conclusions}
In this paper, we have used a game-theoretic to study power control for a wireless data 
network in frequency-selective environments, where the user terminals transmit IR-UWB 
signals and the common concentration point employs ARake receivers. A noncooperative 
game has been proposed in which users are allowed to choose their 
transmit powers according to a utility-maximizing criterion, where the utility function 
has been defined as the ratio of the overall throughput to the transmit power. For this 
utility function, we have shown that there exists a unique Nash equilibrium for the 
proposed game, but, due to the frequency selective multipath, this equilibrium is achieved 
at a different output SINR for each user, depending on the channel realization. 
Using a large system analysis, we have obtained explicit expressions for the utilities 
achieved at the equilibrium. It has also been shown that, under certain conditions, the 
noncooperative solution leads to a nearly SINR-balancing scenario. In order to evaluate 
the efficiency of the Nash equilibrium, we have studied an optimum cooperative solution, 
where the network seeks to maximize the sum of the users' utilities. It has been shown that 
the difference in performance between Nash and cooperative solutions is not 
significant for typical values of network parameters.

Further improvement in the proposed analysis can be achieved using additional mathematical
tools, including the weak version of the law of large numbers. This approach allows a
more accurate approximation of the interfering terms to be derived. Furthermore, a more
general model can be described, which considers different types of Rake receivers and a
broader class of power delay profiles.


\begin{thebibliography}{99}


\bibitem{mackenzie}
A.B.~MacKenzie and S.B.~Wicker, ``Game theory in communications:
Motivation, explanation, and application to power control,''
in \emph{Proc. IEEE Globecom Telecommun. Conf.}, 
San Antonio, TX, 2001, pp. 821-826.

\bibitem{goodman1}
D.J.~Goodman and N.B.~Mandayam, ``Power control for wireless
data,'' \emph{IEEE Pers. Commun.}, Vol. 7, pp. 48-54, Apr. 2000.

\bibitem{goodman2}
D.J.~Goodman and N.B.~Mandayam, ``Network assisted power control
for wireless data,'' in \emph{Proc. IEEE Veh. Technol. Conf.}, Rhodes, Greece,
2001, pp. 1022-1026.

\bibitem{saraydar1}
C.U.~Saraydar, N.B.~Mandayam and D.J.~Goodman, ``Pricing and
power control in a multicell wireless data network,'' \emph{IEEE J. Sel. Areas
Commun.}, Vol. 19 (10), pp. 1883-1892, Oct. 2001.

\bibitem{saraydar2}
C.U.~Saraydar, N.B.~Mandayam and D.J.~Goodman, ``Efficient power
control via pricing in wireless data networks,'' \emph{IEEE Trans. Commun.},
Vol. 50 (2), pp. 291-303, Feb. 2002.

\bibitem{feng}
N.~Feng, S.-C.~Mau and N.B.~Mandayam, ``Pricing and power control
for joint network-centric and user-centric radio resource management,''
\emph{IEEE Trans. Commun.}, Vol. 52 (9), pp. 1547-1557, Sep. 2004.

\begin{figure}
  \centering
  \includegraphics[width=8.85cm]{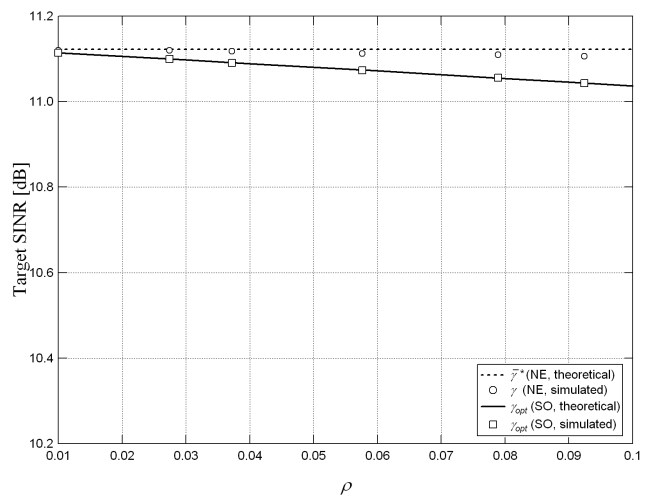}
  \caption{Comparison of the target SINRs versus load factor for the noncooperative and 
    Pareto-optimal solutions.}
  \label{fig:gammaNEvsSO}
\end{figure}

\bibitem{meshkati1}
F.~Meshkati, M.~Chiang, H.V.~Poor and S.C.~Schwartz, ``A
game-theoretic approach to energy-efficient power control in
multicarrier CDMA systems,'' \emph{IEEE J. Sel. Areas Commun.}, Vol. 24
(6), pp. 1115-1129, Jun. 2006.

\bibitem{meshkati2}
F.~Meshkati, H.V.~Poor, S.C.~Schwartz and N.B.~Mandayam, ``An
energy-efficient approach to power control and receiver design in
wireless data networks,'' \emph{IEEE Trans. Commun.}, Vol. 53 (11),
pp. 1885-1894, Nov. 2005.


\bibitem{win}
M.Z.~Win and R.A.~Scholtz, ``Ultra-wide band time-hopping
spread-spectrum impulse radio for wireless multi-access
communications,'' \emph{IEEE Trans. Commun.}, Vol. 48 (4), pp. 679-691, Apr. 2000.


\bibitem{nakache}
Y.-P.~Nakache and A.F.~Molisch, ``Spectral shape of UWB signals
influence of modulation format, multiple access scheme, and pulse
shape,'' in \emph{Proc. IEEE Veh. Technol. Conf.}, Jeju, Korea, 2003, pp. 2510-2514.

\bibitem{fishler1}
E.~Fishler and H.V.~Poor, ``On the tradeoff between two types
of processing gains,'' \emph{IEEE Trans. Commun.}, Vol. 53 (10), Oct. 2005,
pp. 1744-1753.

\bibitem{fcc}
U.S. Federal Communications Commission, FCC 02-48: First Report and Order.

\bibitem{molisch}
A.F.~Molisch, J.R.~Foerster and M.~Pendergrass, ``Channel models for
ultrawideband personal area networks,'' \emph{IEEE Wireless Commun.}, Vol. 10
(6), pp. 14-21, Dec. 2003.

\bibitem{gezici}
S.~Gezici, H.~Kobayashi, H.V.~Poor and A.F.~Molisch, ``Performance
evaluation of impulse radio UWB systems with pulse-based polarity
randomization,'' \emph{IEEE Trans. Signal Process.}, Vol. 53 (7), pp.
2537-2549, Jul. 2005.

\bibitem{hayajneh}
M.~Hayajneh and C.T.~Abdallah, ``Statistical learning theory to
evaluate the performance of game theoretic power control algorithms
for wireless data in arbitrary channels,'' in \emph{Proc. IEEE Wireless Commun. 
and Networking Conf.}, New Orleans, LA, 2003, pp. 723-728.

\bibitem{sun}
J.~Sun and E.~Modiano, ``Opportunistic power allocation for
fading channels with non-cooperative users and random access,''
in \emph{Proc. IEEE Int. Conf. on Broadband Networks}, Boston, MA, 2005, 
pp. 366-374.

\bibitem{huang}
M.~Huang, P.E.~Caines and R.P.~Malham{\'e}, ``Individual and mass
behaviour in large population stochastic wireless power control
problems: centralized and Nash equilibrium solutions,'' in \emph{Proc. IEEE 
Conf. on Decision and Control}, Maui, HI, 2003, pp. 98-103.

\bibitem{rodriguez}
V.~Rodriguez, ``An analytical foundation for resource management
in wireless communications,'' in \emph{Proc. IEEE Globecom Telecommun. Conf.},
San Francisco, CA, 2003, pp. 898-902.


\bibitem{fishler}
E.~Fishler and H.V.~Poor, ``Low-complexity multiuser detectors for time-hopping 
impulse-radio systems,'' \emph{IEEE Trans. Signal Process.}, Vol. 50 (9), pp. 1440-1450, 
Sep. 2002.
\bibitem{verdu}
S.~Verd{\'u}, \emph{Multiuser Detection}. Cambridge, UK: Cambridge Univ. Press, 1998.
\bibitem{proakis}
J.G.~Proakis, \emph{Digital Communications}, 4th ed. NY: McGraw-Hill, 2001.

\bibitem{bacci}
G.~Bacci, M.~Luise, H.V.~Poor and A.M.~Tulino, ``Energy-efficient power control in
impulse radio UWB wireless networks,'' preprint, Princeton University. 
[Online]. Available: http://arxiv.org/PS\verb,_,cache/cs/pdf/0701/0701017.pdf.

\bibitem{rappaport}
T.S.~Rappaport, \emph{Wireless Communications: Principles and Practice,} 2nd ed. 
NJ: Prentice-Hall, 2001.

\bibitem{cassioli}
D.~Cassioli, M.Z.~Win and A.F.~Molisch, ``The ultra-wide bandwidth
indoor channel: From statistical model to simulations,'' \emph{IEEE J. Sel.
Areas Commun.}, Vol. 20 (6), pp. 1247-1257, Aug. 2002.


\end{thebibliography}
\end{document}